# Novel Observation of Piezoelectricity in VO$_2$

Raktima Basu[1, 2, *], G Mangamma[1, *], and Sandip Dhara[1]

[1] Surface and Nanoscience Division, Indira Gandhi Centre for Atomic Research, Homi Bhabha National Institute, Kalpakkam-603102, India

[2] Presently at Department of Physical Sciences, Indian Institute of Science Education and Research, Kolkata, Mohanpur Campus, Mohanpur-741246, Nadia, West Bengal, India

*Email: raktimabasu14@gmail.com ; gm@igcar.gov.in

*Abstract*

VO$_2$ is well known for its dual phase transitions; electrical as well as structural, at a single temperature of 340K. The low temperature structural phases of VO$_2$ are different from its high temperature counterpart by means of structural symmetry. The strain induced modification of the structural distortion in VO$_2$ is studied in details. A ferroelectric type distortion is observed, and therefore, the piezoelectric effect in the low temperature phases of VO$_2$ is investigated, for the first time, using piezo-response force microscopy. The electronic behavior of piezoelectric materials can be tuned with the application of mechanical strain and strain is the only factor to separate the two low-temperature phases, namely, M1 and M2 in the phase diagram of VO$_2$. The piezo-electric coefficient in the strained phase of VO$_2$ was found as 11-12 pm/V making it eligible for piezotronic applications.

## Introduction

Piezoelectric materials find current attention because of their fruitful applications in renewable energy production, such as self-powered devices, micro-actuators, pressure-sensors, ultrasonic motors, mechanical energy harvesting among others.[1–5] Piezoelectricity is nothing but production of electrical potential inside a non-centrosymmetric material subjected to a mechanical strain or vice versa. Therefore, all ferroelectric materials exhibit piezoelectric effect due to lack of symmetry. Among vanadium oxides, $V_2O_5$ is reported to be a ferroelectric material with Curie temperature of 530K.[6] The piezoelectricity, and thus ferroelectricity in $VO_2$ has not been investigated so far, although predicted in previous reports.[7] However, $VO_2$ draws significant attention for its well-known metal to insulator transition (MIT) at a technologically important temperature of 340K, which is very close to room temperature.[8] $VO_2$ crystallizes in rutile tetragonal (R; space group $P4_2/mnm$) and monoclinic (M1; space group $P2_1/c$) structure above and below the transition temperature, respectively.[9,10] In the high-temperature R phase, V atoms are equally spaced, forming linear chains along the $c_R$ axis with each V atom surrounded by an oxygen octahedron (Fig. 1a).[11] The lattice parameters are $c_R$ = 2.85 Å, and $a_R = b_R$ = 4.55 Å. Whereas, in the low-temperature monoclinic phase the volume of the unit cell becomes double than that of R-phase with lattice parameters $a_{M1}$ = 5.70 Å, $b_{M1}$ = 4.55 Å, $c_{M1}$ = 5.38 Å, and $ß_{M1}$ = 123°.[12] In the M1-phase there are significant differences in the arrangement of V atoms along $c_R$ axis. The V atoms form pair, and the pairs tilt along the $c_R$ axis, which leads the vanadium ion moving away from the center of the oxide octahedron (Fig. 1b). The deformation of the octahedron changes the identical V-O bonds of R-phase to different long and short V-O bonds.[13] Besides M1, another monoclinic phases of $VO_2$, M2 (space group $C2/m$) is also reported to evolve during the phase transition.[14] The metastable M2-phase of $VO_2$ is reported to be stabilized at room temperature by introducing strain in the system; either via mechanical strain or via doping with metals of lower valency than $V^{4+}$ (e.g., $Al^{+3}$, $Ga^{+3}$, $Cr^{+3}$).[15,16] For M2-phase, in one of the sublattices, the V ions along the $c_R$ axis dimerized without twisting, while the V ions in the nearest sub-lattice remain non-dimerized and form canted V-V chains (Fig. 1c). Due to different arrangements of V atoms in subsequent sublattice, the oxide octahedra deform from their regular shape as in R phase. The lattice parameters for M2-phase are: $a_{M2}$ = 9.07 Å, $b_{M2}$ = 5.797 Å, $c_{M2}$ = 4.53 Å, and $ß_{M2}$ = 91.88°.[17]

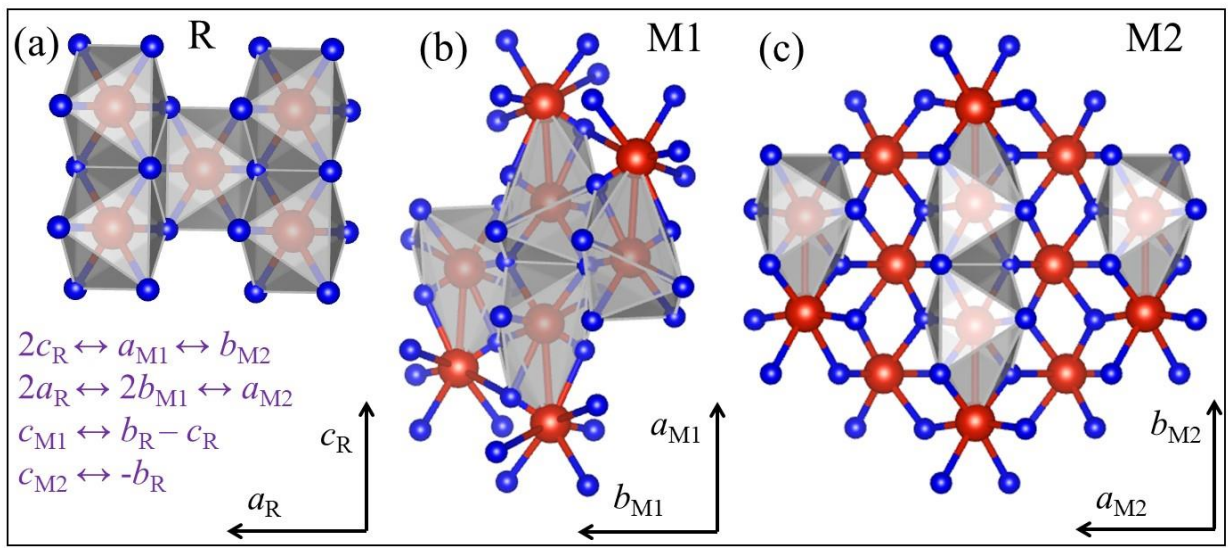

**Fig. 1.** The schematic structures for (a) rutile tetragonal R, (b) monoclinic, M1 and (c) monoclinic, M2 phases of $VO_2$. Blue and brown balls denote V and O atoms, respectively.

In both the low-temperature phases of $VO_2$, the tilting of $c_R$-axis shortens the V-O separation perpendicular to the $c_R$-axis.[14] The displacement of a cation from the center of its interstice toward one or more anions is characteristic of a ferroelectric-type distortion. The dielectric constant of $VO_2$ is also reported to be low (~40),[18] and so is of special interest for piezoelectric effect. Moreover, the electronic behavior of piezoelectric materials can be tuned with the application of mechanical strain and strain is the only factor to separate the two low-temperature phases (M1 and M2) in the phase diagram of $VO_2$.[19,20] In the present report, therefore, we have studied the piezoelectric effect in both M1 and M2 phase of $VO_2$, for the first time, using piezo-response force microscopy (PFM).

**Experimental details**

$VO_2$ micro-crystals were synthesized on Si(111) substrate by vapor transport process using Ar as carrier gas. Bulk $VO_2$ powder (Sigma-Aldrich, 99%) was used as source and was placed in a high pure (99.99%) alumina boat inside a quartz tube. The synthesis was carried out at 1150K for 3h. Sample S2 was prepared in presence of Mg dopant, whereas, in case of sample S1, no dopant was used. Structural properties of the as-grown samples were studied using x-ray diffraction in beamline-11, installed at Indus-2 synchrotron facility at RRCAT, India, with λ = 0.76089 Å using Si(111) channel-cut monochromator. Calibration and conversion of 2-D diffraction data to 1-D, intensity vs. 2θ, were carried out using the FIT2D software.[21] Rietveld refinement was carried out using GSAS + EXPGUI.[22] We have used VESTA[23] software for visualization of the crystal structure and calculation of the bond lengths after Rietveld refinement. The vibrational modes of the grown sample were studied by Raman spectroscopic analysis using a micro-Raman spectrometer (inVia, Renishaw, UK) in the backscattering configuration with Ar+ Laser (514.5 nm) as excitation source, diffraction gratings of 1800 gr.mm$^{-1}$ and a thermoelectrically cooled CCD camera as the detector.

The piezoelectric response of VO$_2$ microcrystals were measured by PFM configuration (NT- MDT, NTEGRA) in contact mode with an electrically conductive atomic force microscopic tip. An AC bias, [$V_{ac}\cos(\omega t)$] was applied between the cantilever and the sample surface to activate the domains. The amplitude of the PFM response reflects the strength of the piezo-response, and the phase signal provides information about the polarization direction of the sample. An external DC bias (ranging from −3V to +3V) and AC bias with amplitude, 0.3V and frequency, 255 kHz was applied between the tip and the sample. For the measurement of $d_{33}$, we used PFM spectroscopy for $V_{dc}$ = 0 with minimum $V_{ac}$ value to reduce contribution of the electrostatic term. A diamond-like-carbon coated stiff cantilever of length, 100 μm; width, 35 μm; and thickness, 2 μm and the spring constant of 11 N/m was used as tip. The lock-in amplifier de-convolutes the induced signal on the cantilever to measure the amplitude and phase difference with respect to the input ac-voltage and produce the respective images over the defined scanned area.

## Results and discussions

The FESEM images of the pristine samples are shown in figure 2. The as-grown microcrystals have average width of 2 to 5 μm for sample S1 (fig. 2a). The insets of figure 2a show magnified image of a single crystal microrod of width ~2±0.5 μm. Whereas in case of sample S2 (fig. 2b) both micro and nano-rods were observed to be present. The average width of microrods is of 2 to 6 μm and that for nanorods are of ~200±40 nm for sample S2. The (011) plane of monoclinic M1 phase is the preferential growth plane for VO$_2$. Our earlier HRTEM studies for a single crystal VO$_2$ showed the sample in (011) orientation and growth axis along the [100]$_{M1}$.[24]

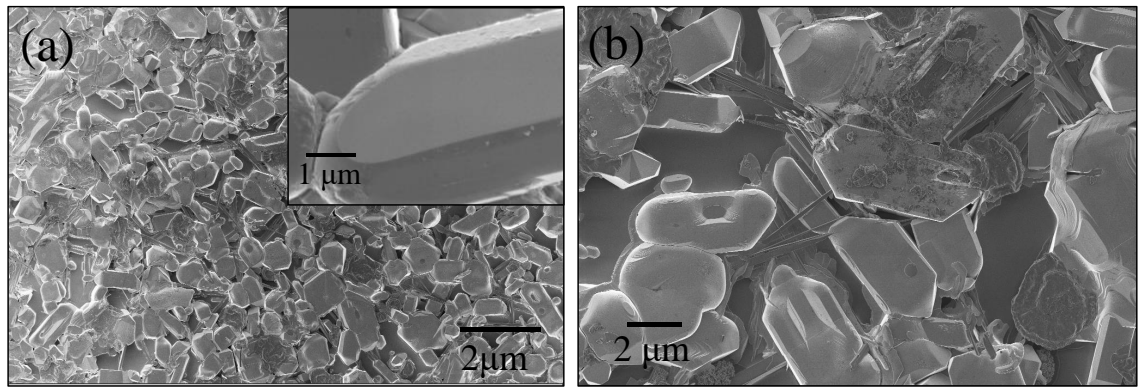

**Fig. 2.** FESEM images of as-grown micro-crystals for (a) sample S1 and micro and nano-crystals for (b) sample S2. Inset of (a) shows magnified image of a single microrod.

The crystallographic orientation and phase confirmation of the pristine samples were studied by the Rietveld refined x-ray crystallographic patterns as shown in figure 3a-b. In sample S1 (Fig. 3a), the diffraction peaks match with the monoclinic M1 phase of $VO_2$ (JCPDS # 04-007-1466)[25] with $R_{WP}$=0.056 $R_p$=0.043. Whereas the diffraction peaks for sample S2 (Fig. 3b) confirm the presence of M2 phase of $VO_2$ (JCPDS # 01-071-0289)[26] with $R_{WP}$ = 0.048 $R_p$ = 0.036. While focusing on the diffraction peaks at lower 2θ values, the diffraction peak at 2θ=13.67° represents the (011) plane corresponding to monoclinic M1 phase (equivalent to $(110)_R$ plane) of $VO_2$ for samples S1. In the case of sample S2, two diffraction peaks observed at 2θ = 13.38° and 13.84° correspond to (-201) and (201) planes of M2 phase of $VO_2$.[27]

The M2 phase is reported as the strained (tensile strain along the rutile $c_R$ axis) stage for the M1 phase.[17,19] In the present study, the role of substrate in introducing strain in these samples is ruled out as they all are synthesized on the same substrate. Mg as dopant introduces strain in the sample and helps in stabilizing the M2 phases of $VO_2$, as reported in our previous studies.[28,29] The lattice parameter $b_{M2} > a_{M1}$ (equivalent to $c_R$) indicates a tensile strain along $c_R$ axis, which may be responsible for stabilizing the M2 phase. The tensile strain along the $c_R$ axis is found to be ~8.2×10⁻³.

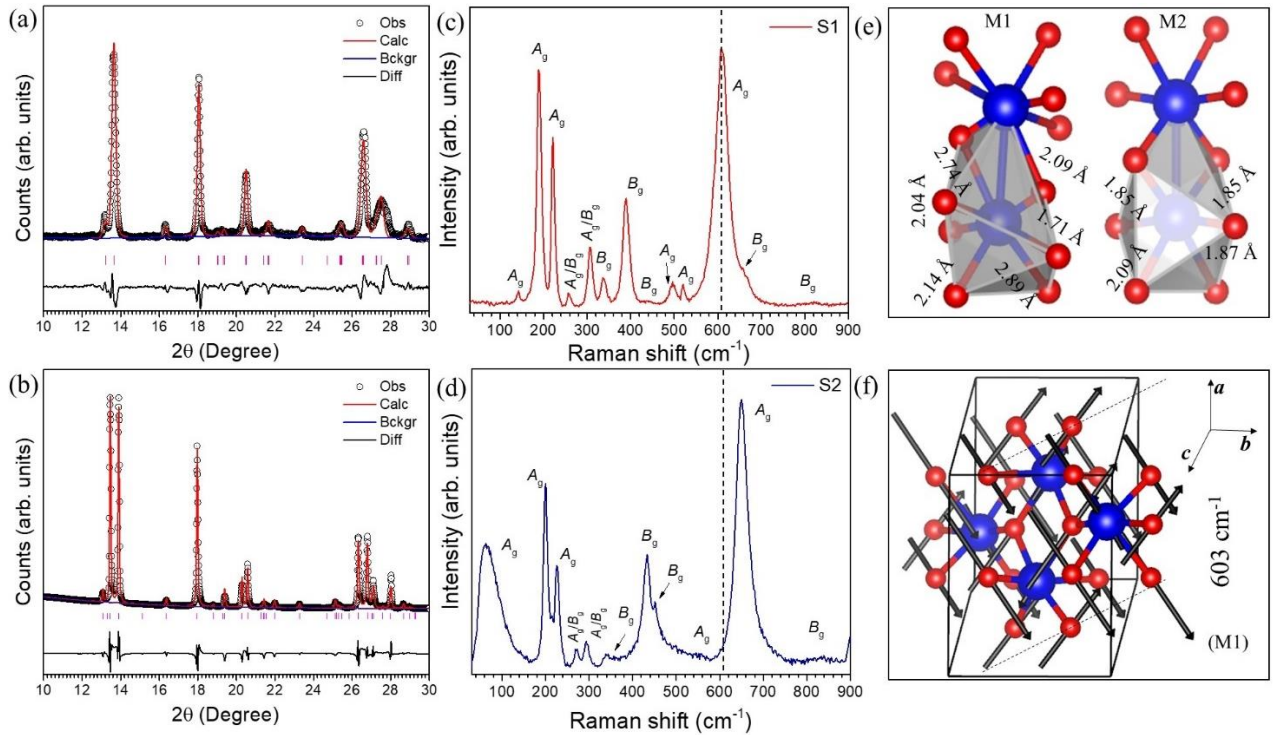

**Fig. 3.** Rietveld fitted diffraction pattern for samples (a) S1 and (b) S2. Raman spectra with proper symmetry notations of the pristine samples (c) S1, and (d) S2. Dashed lines are guide to eye. (e) Schematic of V-O bond lengths of M1 and M2 calculated using VESTA, (f) Schematic atomic displacement for the 603 cm⁻¹ phonon mode of M1 phase. The directions of the displacements of atoms are shown by the arrows. The V and O atoms are shown with large (blue) and small (red) balls, respectively.

We calculated the V-V and V-O bond lengths after Rietveld refinement using VESTA software. The V-V lengths were found as 2.60 (intra-dimer) and 3.11 Å (inter-dimer) along $c_R$ axis. Whereas, in case of M2 phase, V-V separations are found out as 2.51 Å (intra-dimer) and 3.26 Å (inter-dimer) for the dimerized V-chains along $c_R$ axis. However, for the non-dimerized zig-zag V- chains all V-V bonds were found to have a single length of 2.95 Å. The observed bond lengths are almost similar as reported earlier.[7,30] The V-O distances in the deformed octahedron are found out as 2.89, and 2.74 (bridging oxygens between paired vanadium ions); 2.04, and 2.09 (connecting oxygens between two V-chains); 2.14, and 1.71 Å (Fig. 3e) for M1 phase. Whereas, for M2 phase (Fig. 3e), the V-O separations between the V- chains i.e., perpendicular to (011) plane are found to be shortened (1.85 and 1.87 Å). The tensile strain along $c_R$ axis may lead to compression in its perpendicular direction. The other pairs of V-O bond-length is found to be 2.09 Å.

Raman spectroscopic analysis was carried out to obtain further information about the phases present in the as-grown samples. The Raman spectra for the samples collected at room temperature are depicted in figure 3c-d. Eighteen Raman-active phonon modes are predicted by Group theoretical analysis for all low-temperature phases of $VO_2$ (for M1: $9A_g+9B_g$, and for M2: $10A_g+8B_g$) at $\Gamma$ point with different symmetries.[31,32]

However, we observed twelve vibrational modes for samples S1 (Fig. 3c). Observed Raman modes at 141, 190($A_g$), 225($A_g$), 258 (either $A_g$ or $B_g$,; $A_g/B_g$), 307($A_g/B_g$), 335($A_g$), 390($A_g/B_g$), 440($A_g/B_g$), 497($A_g/B_g$), 609($A_g$), 665($B_g$), 823($B_g$) cm$^{-1}$ in case of sample S1 (Fig. 3c) confirms the presence of pure M1 phase of $VO_2$.[33,34] However, for samples S2, we observed eleven Raman modes at ~ 50, 203 ($A_g$), 217($A_g$), 229($A_g$), 273($A_g/B_g$), 297($A_g$), 341($A_g$), 432($A_g/B_g$), 454($A_g/B_g$), 651($A_g$), and 831($B_g$) cm$^{-1}$ (Fig. 3d), which exactly resembled the reported M2 phase of $VO_2$.[15,35]

The Raman mode, observed at 609 cm$^{-1}$ in sample S1, is reported due to V-O stretching vibration.[32] We have carried out the density functional theory (DFT) calculations for the phonon density of states. The detail of the study is published in one of our previous works.[24] We observed phonon mode at 609 cm$^{-1}$ corresponds to the calculated mode at 603 cm$^{-1}$ as shown schematically in figure 3f for the atomic displacements. The V-O vibrations perpendicular to the (011) plane contributes maximum for the Raman mode frequency at 609 cm$^{-1}$, which shifts to 651 cm$^{-1}$ in sample S2 (as indicated by dashed line in Fig. 3c-d). The blue shift of the Raman mode indicates that the length of V-O bond became shorter with Mg doping. In pure $VO_2$, $V^{4+}$ ion is positioned at the center of the octahedron constituting of oxygen atoms, and the principal axes of the octahedron are directed perpendicular to $(011)_{M1}$ lattice plane.[16] Each V atom shared its four electrons with six neighbor O atoms, and each O atom attracted adjacent electrons supplying by three nearest V atoms. After $Mg^{2+}$ ion occupies the native $V^{4+}$ ($d^1$) place, the adjacent $V^{4+}$ ($d^1$) sites in the neighboring chains got replaced by $V^{+5}$ ($d^0$) sites.[29,36] The replacement of $V^{4+}$ by $V^{5+}$ leads two apical $O^{2-}$ of the octahedron move closer to each other resulting reduction in the V-O bond length,[37] as also observed from XRD analysis. Figure 3e shows that the oxide-octrahedra in both M1 and M2 phase got distorted and assymetric due to off-centre displacement of the central V ion. The assymetry gives rise to a resultant dipole to the crystals along the priciple axes of the octahedra (perpendicular to (011)$_{M1}$ plane). As the structures of both the samples shows ferroelectric type distortion, we have carried out the peizoelectric studies on both the samples.

The piezo-response imaging was carried out with the application of combined DC and AC voltage applied between the tip and the sample. One of the typical topographies, magnitude and phase images collected over an area of 50 × 50 μm² is presented in fig. 4a-c for sample S1 and fig. 4d-f for sample S2. In the topographic image of S1 (Fig. 4a), the microcrystals of size 2-5 are observed to be distributed over the area. In case of sample S2, micro crystals as well as nano-crystals (Fig. 4b) are observed as shown in FESEM images (Fig. 2).

The magnitude and phase images of sample S1 do not show considerable variation in the contrast as compared to the background, signifying very less piezo-response for sample S1 (Fig. 4b-c). However, in case of sample S2, the magnitude and phase images (Fig. 4e-f) show the bright contrast with respect to the substrate signifying the piezoelectric deformation in the probed direction. Moreover, the bright contrast implies that the applied field and the polarization direction are parallel (**E**∥**P**) to each other.

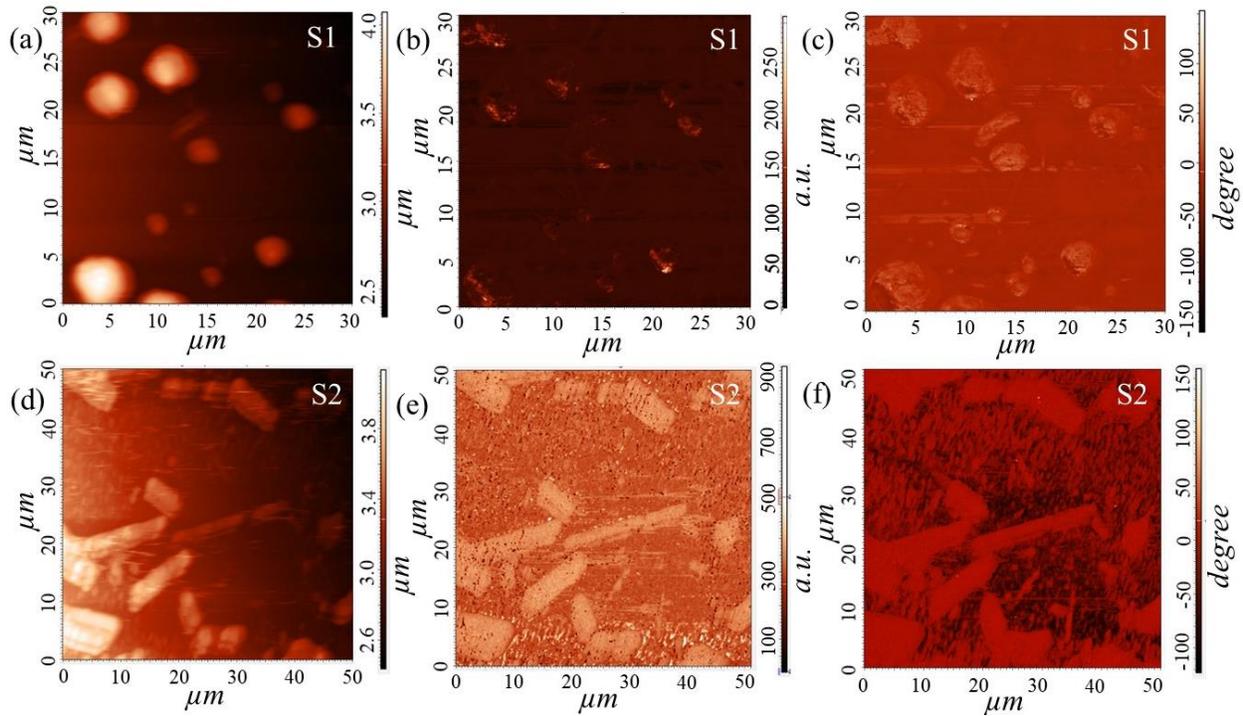

**Fig. 4.** (a) The topographic, (b) magnitude, and (c) phase images of sample S1. Corresponding (d) topography, (e) magnitude, and (f) phase images of sample S2.

As, the samples are oriented by (011)$_{M1}$ face, the electric field direction and the principle axis of the octrahedra are parallel to each other resulting in bright contrast in the piezo-response imaging.

The electronic behavior of VO$_2$ is reported to be tuned with application of strain, which is also true for piezoelectric materials. The M2 phase is confirmed as the strained version of M1 phase of VO$_2$. The cation-anion (V-O) distance along the polar direction is also found to be reduced in M2 phase. The weak piezo-response in M1 phase might be modified with the induced strain, leading M2 phase as ferroelectric.

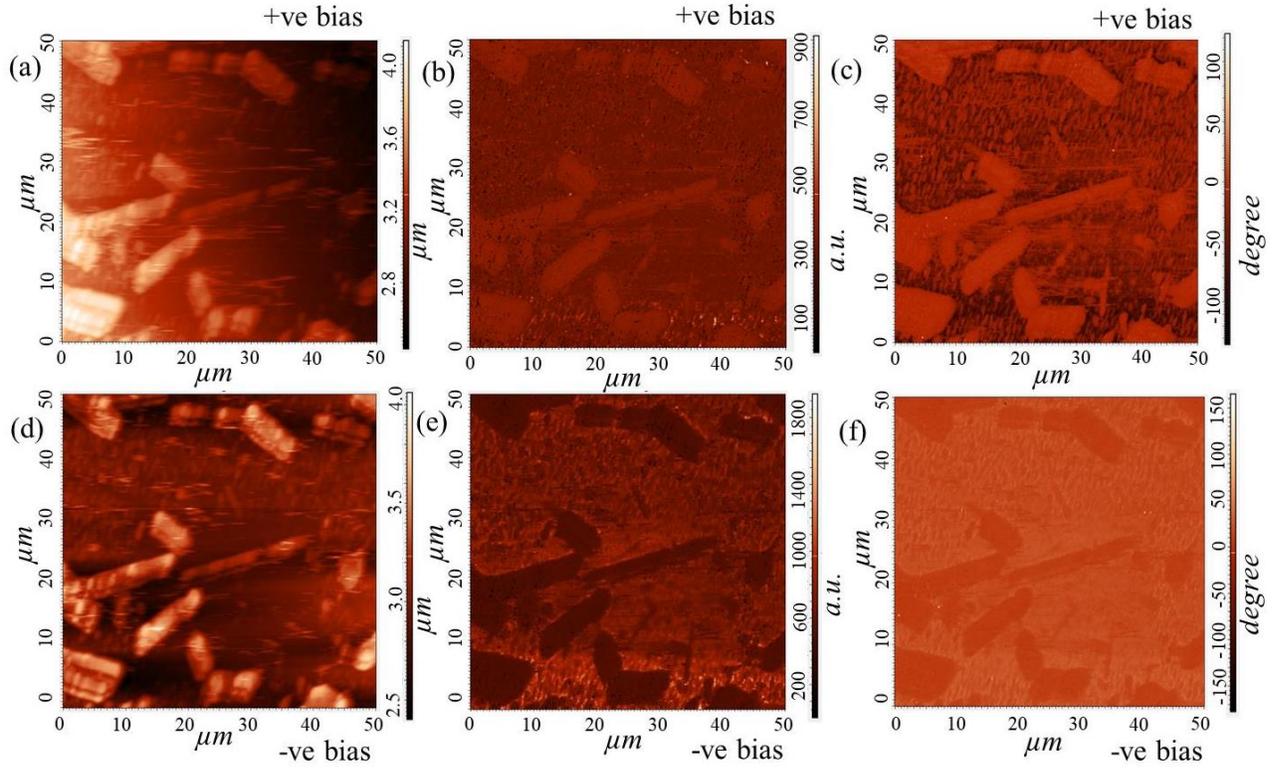

**Fig. 5.** (a) The topographic, (b) magnitude, and (c) phase images of sample S2 at +3V. Corresponding (d) topography, (e) magnitude, and (f) phase images at -3V.

Further to enrich our argument, the PFM spectroscopy was carried out on several sets of alternative bias voltages. Figure 5 shows the typical topography, magnitude and phase images collected at bias of +3V and -3V. While applying +ve bias, the phase image (Fig. 5c) shows bright contrast with respect to the substrate, whereas, upon application of –ve bias, the sample shows dark contrast (Fig. 5f) with respect to the substrate, indicating the piezoelectric nature the sample.

To support our claim, we have carried out the PFM spectroscopy on sample S1 and S2 (M1, and M2 phase of VO$_2$, respectively) in the applied voltage range of −10 to +10 V in forward and backward sweep. The topography images are also shown in fig. 6 along with the change in magnitude and phase over the voltage range.

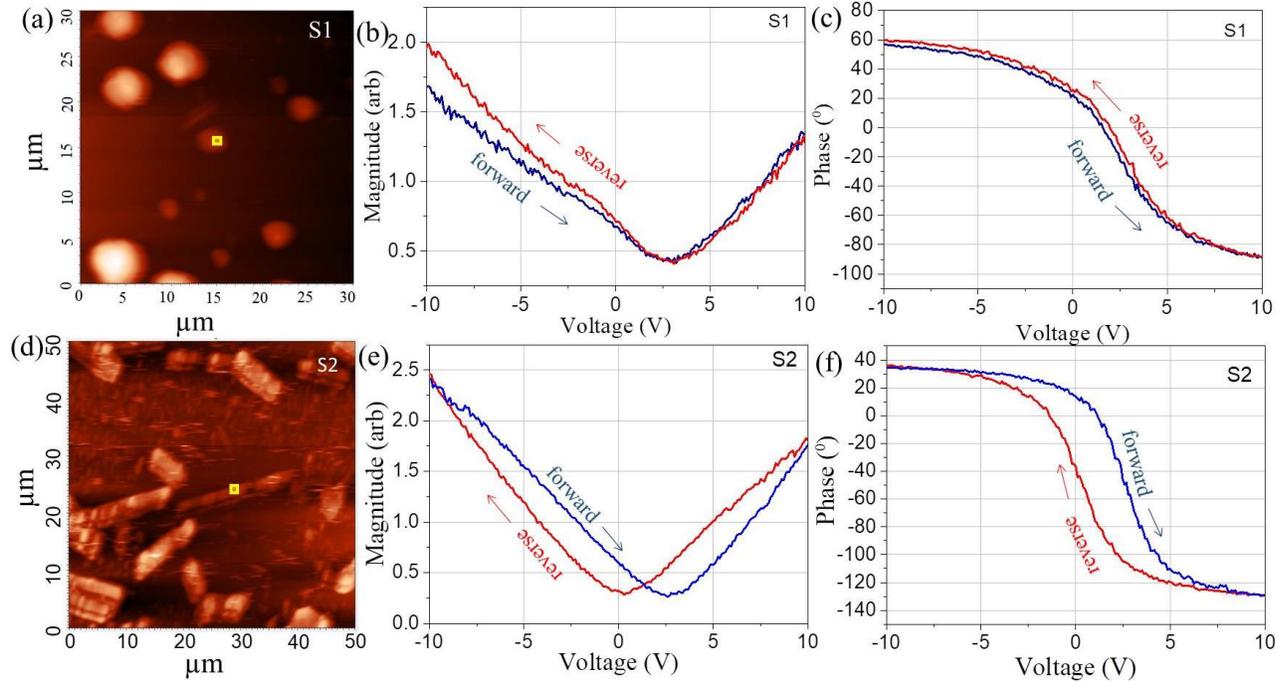

**Fig. 6.** (a) The topography, change in (b) magnitude, and (c) phase of sample S1 in the applied voltage range of −10 to +10 V. Corresponding (d) topography, change in (e) magnitude, and (f) phase of sample S2 in the applied voltage range of −10 to +10 V.

In both positive and negative sweep, the piezo-response increases with increase in magnitude of the applied voltage for both the samples. However, sample S1 does not show any hysteresis loop both in magnitude (Fig. 6b) and phase (Fig. 6c) cycle over the entire voltage range, suggesting the sample as non-ferroelectric. Figures 5e-f show hysteresis towards the piezo-response with the voltage sweep for both the magnitude and phase cycles for sample, S2. This hysteresis signifies that the observed piezo-response is because of the induced polarization added up with the spontaneous polarization arising from the sample structure confirming M2 as ferroelectric material too.

The hysteresis curves (magnitude and phase loops) for sample S2 show perfect shape apart from 2-2.5 V shift with complete saturation signifying no leakage-current, which makes it applicable as functional device. The similar voltage shift is observed for sample S1 as well. The shift of local PFM hysteresis loops from origin may be due to the presence of a biased-voltage generated between different contacts of the bottom and top electrodes to the sample.[38]

A material to be ferroelectric its structures should be such that the correlation of the distortions of neighboring octahedra to give a resultant dipole to the crystal.[7] In case of M1 phase, though the oxide octahedron is asymmetric enough, along the polar direction the net dipole moment is zero due to anti-ferroelectric type distortion (Fig. 1b). However, in case of M2 phase, due to induced strain the atomic distribution of V along the $c_R$ axis is different in two sub lattices (Fig. 1c). Two different types of distortion in the neighboring V chains gives rise to a net dipole moment along the polar plane, making M2 as ferroelectric.

To quantify the observed piezo-response, the $d_{33}$ value was calculated using the point spectroscopy for $V_{dc}$ = 0 V. In a typical PFM, when a modulation voltage V is applied between the tip and the piezoelectric material, there is a vertical displacement of the tip. The tip accurately follows the piezoelectric motion of the sample surface due to mechanical contact with the sample. In our study, the voltage applied across the microcrystal of height $h$, generates an electric field $E_3$ along the $c$-axis, which elongates or contracts the crystal by an amount $\Delta h$. Then the piezoelectric coefficient $d_{33}$ is given as,[39]

$$d_{33} = \Delta S_3/E_3 \quad \ldots\ldots\ldots(1)$$

Where $E_3 = V/h$ and $\Delta S_3 = \Delta h/h$ is the change of strain along the $c$-axis. Putting the values of $E_3$ and $\Delta S_3$ in Eq. (1), $d_{33}$ can be written as

$$d_{33} = \Delta h/V \quad \ldots\ldots\ldots(2)$$

Thus, the amplitude of the tip vibration provides information on the piezoelectric strain, and the piezoelectric coefficient, $d_{33}$ can be determined using Eq. (2). In the present case, the magnitude signal was obtained in pA. With the help of the force-distance curve, the signal in pA is converted to corresponding value in pm. $d_{33}$ value is calculated by dividing the piezoelectric deformation with the corresponding AC voltage. The average value of the $d_{33}$ was calculated from the different micro and nano-crystals and found to be as 11-12 pm/V (with <10% error). The value of $d_{33}$ is quite good in comparison to other semiconducting piezo-materials such as ZnO,[40] ZnS,[41] and III-V nitrides.[42-44] Our study confirms $VO_2$, which is well known for its electronic and optoelectronic properties, is also a suitable piezo-material in its M2 phase, for the first time, making it eligible for piezotronic applications.

## Conclusions

The $VO_2$ micro and nanocrystals, in two different structural phases (M1 and M2), are synthesized by vapour transport technique. The M2 phase was found to be stabilized by introduction of strain via Mg doping. The piezo-response force microscopic imaging was carried on both the pristine phases. The $VO_2$ micro- and nanocrystals, stabilized in M2 phase, showed piezoelectric deformation because of the spontaneous polarization. Whereas, the $VO_2$ microcrystals grown in M1 phase, show very weak piezo-response to the applied field confirming absence of any spontaneous polarization. The induced strain in the M2 phase is anticipated as the main reason for the piezo-response. M2 phase was also explored as ferroelectric, for the first time, due to presence of net dipole moment in the neighboring distorted octahedra and spontaneous polarization along probed direction. A piezoelectric coefficient, $d_{33}$ ~11-12 pm/V for the M2 phase of $VO_2$, is reported for the first time using the PFM technique. The present study provides insight in VO$_2$, which is well known for its electronic and optoelectronic properties, is also eligible for piezotronic applications.

## Acknowledgments

We acknowledge R. Pandian of SND, IGCAR for FESEM studies, and Sharat Chandra, MPD, IGCAR for DFT calculation. We also thank V. Srihari of SRPD, BARC, Mumbai, for XRD studies.

## Notes and references